\documentclass[]{spie}  

\usepackage{amsmath,amsfonts,amssymb}
\usepackage{graphicx}
\usepackage[colorlinks=true, allcolors=blue]{hyperref}

\title{Efficient Brain Extraction of MRI Scans with Mild to Moderate Neuropathology}

\author[a]{Hjalti Thrastarson}
\author[a]{Lotta M. Ellingsen}
\author[*]{for the Alzheimer's Disease Neuroimaging Initiative and the ASAP Neuroimaging Initiative}
\affil[a]{Faculty of Electrical and Computer Engineering, University of Iceland, Reykjavík, Iceland}
\renewcommand{\sup}[1]{}

\authorinfo{Further author information: Send correspondence to H.Th: E-mail: hth262@hi.is or L. M. E.: E-mail: lotta@hi.is } 

\begin{document} 
\maketitle

\begin{abstract}
Skull stripping magnetic resonance images (MRI) of the human brain is an important process in many image processing techniques, such as automatic segmentation of brain structures. Numerous methods have been developed to perform this task, however, they often fail in the presence of neuropathology and can be inconsistent in defining the boundary of the brain mask. Here, we propose a novel approach to skull strip T1-weighted images in a robust and efficient manner, aiming to consistently segment the outer surface of the brain, including the sulcal cerebrospinal fluid (CSF), while excluding the full extent of the subarachnoid space and meninges. We train a modified version of the U-net on silver-standard ground truth data using a novel loss function based on the signed-distance transform (SDT). We validate our model both qualitatively and quantitatively using held-out data from the training dataset, as well as an independent external dataset. The brain masks used for evaluation partially or fully include the subarachnoid space, which may introduce bias into the comparison; nonetheless, our model demonstrates strong performance on the held-out test data, achieving a consistent mean Dice similarity coefficient (DSC) of 0.964$\pm$0.006 and an average symmetric surface distance (ASSD) of 1.4mm$\pm$0.2mm. Performance on the external dataset is comparable, with a DSC of 0.958$\pm$0.006 and an ASSD of 1.7$\pm$0.2mm. Our method achieves performance comparable to or better than existing state-of-the-art methods for brain extraction, particularly in its highly consistent preservation of the brain's outer surface. The method is publicly available on GitHub. 
\end{abstract}


\section{INTRODUCTION}
\label{sec:intro}  

Pre-processing is an essential step in automatic analysis of brain MRIs, with skull stripping (or brain extraction) being one of the most time-consuming and complex pre-processing steps. Numerous automatic skull-stripping methods have been developed over the past decades. The Brain Extraction Tool (BET)\cite{Smith2002} is widely regarded as one of the earliest and most well known methods developed for this task. In later years, other more robust methods have emerged using both classical and neural network-based approaches. One of the classical approaches, Multi-cONtrast Brain Stripping (MONSTR)\cite{Roy2017}, uses an atlas-based patch method to robustly skull strip brains with neuropathology. A highly successful machine learning-based approach is ROBEX\cite{Iglesias2011}, which uses random forest classifiers to segment the brain surface, and then fits a point distribution model to ensure that the mask is plausible. One of the more recent 
methods is Synthstrip\cite{Hoopes2022}, which trains a U-net based model on synthetic data. Numerous other methods have been proposed, however, they all have different pros and cons. 

A key challenge with a lot of the classical methods is that they suffer from long processing times, and even though they are developed to be robust, they can critically fail on subjects with mild to moderate pathology. Deep learning-based methods solve the speed challenge; however, they tend to be attuned to their training datasets. Another challenge with deep learning methods is the need for labeled ground truth data, which can take an experienced rater a long time to manually generate. Finally, a significant challenge common to all these methods is the lack of consensus regarding the inclusion criteria for the brain masks, specifically, whether the subarachnoid space and meninges should be included or not. 

To address these challenges, we propose a skull stripping method developed for MRIs of subjects with mild to moderate neuropathology, aiming to consistently preserve the outer surface of the brain. We present a novel loss function based on the signed distance transform (SDT), building upon the weighted mean square error (MSE) loss function originally proposed by Hoopes et al~\cite{Hoopes2022}. While their method uses cutoff-based weights, we use exponential decay-based weights, which yield a smooth gradient that is defined everywhere and facilitates faster convergence during training. Our method downsamples the images by a factor of 2, enabling larger batch sizes and highly efficient training. Finally, our method requires minimal training data, and can be trained on silver-standard ground truth, which is a significant advantage given the difficulty of obtaining gold-standard annotations. We believe our proposed loss function could benefit other single-object medical image segmentation tasks, where high quality ground truth data is not readily available.

\section{METHOD}

\subsection{Data}

We used two datasets for training and validation. The first comprised 87 heavily standardized T1-weighted MRIs from the Alzheimer's Disease Neuroimaging Initiative (ADNI) (adni.loni.usc.edu). The second dataset was from the curated ASAP Neuroimaging Initative (see acknowledgements), comprising 659 T1-weighted multi-site clinical MRIs of both healthy controls and patients with movement disorders, including Progressive Supranuclear Palsy (PSP), Multiple System Atrophy (MSA) and Parkinson's disease. Due to lack of manual ground truth data, we generated silver-standard ground truth masks by using the STAPLE algorithm\cite{Warfield2004} to combine masks from state-of-the-art methods, thereby producing a probabilistic estimate of the ground truth segmentation. The methods we used to generate these masks were MONSTR, ROBEX, and Synthstrip. Since we aimed to exclude the subarachnoid space outside the sulcal CSF, we included two output masks from Synthstrip, one that included the CSF and another that did not. 

For training we randomly selected 49 images from ADNI and 100 images from ASAP, being careful to swap out images that had significant errors from one or more of the skull stripping methods. We used 14 images from ADNI and 35 from ASAP for validation and held-out the rest for testing. 
For testing on independent external data, we used datasets of T1-weighted images, with both manual and consensus-derived ground truth masks. The authors of Synthstrip created a collection of datasets with consensus masks generated by multiple current skull stripping methods. From this collection we used three datasets with healthy controls for evaluation: 50 MRIs from the IXI dataset from the Biomedical Image Analysis Group, Imperial College London (https://brain-development.org/ixi-dataset), 38 MRIs from the FreeSurfer Maintenance (FSM) dataset from Greve et al.\cite{greve2021deep}, and 43 MRIs from the In-house HCP-A pseudo-continuous arterial spin labeling (ASL) dataset from Harms et al.\cite{harms2018extending}. Additionally, we used 12 manually labeled images from the CC359 datset\cite{souza2018open} for further evaluation. 
For pre-processing, all images were rigidly registered to the Montreal Neurological Institute (MNI) space\cite{Fonov2009}, intensities normalized between 0 and 1, and downsampled by a factor of 2 to battle memory constraints.

\subsection{Model Architecture and Training}

\begin{figure} [ht]
    \begin{center}
    \begin{tabular}{c} 
    \includegraphics[height=8cm]{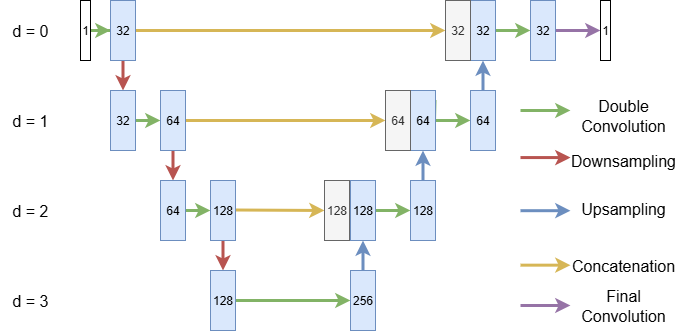}
    \end{tabular}
    \end{center}
    \caption[example] 
       { \label{fig:unet} 
The architecture of the proposed model. The numbers in the blocks signify the number of channels it outputs. Each arrow represents an action, with its color indicating the specific action type.}
   \end{figure}

Our model uses a 3D U-net architecture\cite{Ronneberger2015}, with upsampling performed via transposed convolutions. The model has only two tunable hyperparameters, the depth and the number of starting channels, which control the size and complexity of the model. Figure \ref{fig:unet} shows the model architecture, with the depth set to 3 and the number of starting channels set to 32.

Skull stripping is a specific form of segmentation, and recent studies commonly use the Dice loss\cite{Milletari2016} to train such models. We observed that using Dice loss to train on silver-standard ground truth data gave suboptimal results. Here, we propose a novel loss function based on the SDT, originally suggested by Hoopes et al.\cite{Hoopes2022}. The SDT operates on a binary brain mask by finding its boundary, assigning zero to boundary voxels, and setting other voxel values based on their distance to the boundary, with positive values inside the brain and negative values outside. The original binary mask can then be recovered by thresholding, such that all values greater than or equal to zero are assigned the value one while the rest is set to zero. We formulate a loss function using a weighted MSE loss given by 
\begin{equation}
    L_{SDT}(y, \hat{y}) = \frac{\sum_{i\in\Omega}w_i(y_i - \hat{y}_i)^2}{\sum_{i\in\Omega}w_i},
\end{equation}
\noindent where i represents one voxel in the domain of the image $\Omega$, and w represents the weights of each voxel. The purpose of the weights is to concentrate the gradients around the brain mask boundary, and hence, we want the weights to be high around the boundary and lower deep inside the brain and around the image border. We used exponential decay with an offset for our weights, given by
\begin{equation}
    w_i = e^{-\alpha|\hat{y}|} + \beta.
\end{equation}
\noindent Here, $\alpha$ is a scaling factor and $\beta$ is an offset factor to prevent weights from going to zero. Using the hyperparameter optimization framework Optuna\cite{Akiba2019}, we found that the best values for $\alpha$ and $\beta$ were 0.1 and 0.3, respectively. During training, we also thresholded the validation data to reclaim the masks and calculated the Dice loss for an extra layer of validation, since low SDT loss doesn’t necessarily translate to a low Dice loss. Finally, since the model is trained to estimate the SDT, which is continuous and real-valued with a typical range from -120 to 70, the softmax layer or similar activation function that is typically at the end of a U-net model was omitted. 

The proposed loss function has two key advantages: first, the thresholding step allows adjustable inclusion of brain tissue, with this adjustment only effective near the mask boundaries. Second, it demonstrates greater robustness to inaccuracies in the training data compared to commonly used losses such as the Dice loss, making it particularly well suited for single-object segmentation tasks where properly labelled ground truth data is not readily available. 
For training, we used two Nvidia RTX 4000 Ada Generation GPUs, each with 20 GB of memory. The training was highly efficient, with the model converging in less than 12 hours, primarily due to the use of downsampling. During inference, the images are upsampled and subsequently thresholded, which tends to create a smoothing effect on the resulting brain masks.
For post-processing, all disconnected components were removed, and holes within the brain mask were filled.

\section{RESULTS}

We evaluated our model on held-out subsets from ADNI and ASAP. 
For analysis on independent and previously unseen data, we used 131 images with silver-standard ground truth labels from the  IXI (n=50), FSM (n=38), and ASL (n=43) datasets, as well as 12 images with manual labels from the CC359 dataset. The ground truth in IXI, FSM, and ASL, compiled by the Synthstrip research group, is a majority consensus of multiple state-of-the-art skull stripping methods, with minor manual refinements. It generally includes subarachnoid CSF, while our model does not. Therefore, in addition to quantitative evaluation we perform a qualitative analysis as well. The 12 manually labelled masks from CC359 are conservative in nature and generally do not include any sulcal CSF. Three metrics were used for quantitative analysis: DSC, ASSD, and 95th percentile Hausdorff distance (HD95). 

For the held-out test data, from ADNI and ASAP, we compared our model with other methods using the ground truth masks generated by STAPLE. As seen in Table \ref{tab:heldoutdata}, the proposed method achieved a mean DSC of 0.964$\pm$0.006, mean ASSD of 1.4$\pm$0.2mm, and a mean HD95 of 3.6$\pm$1.1mm. The results for the IXI, FSM, and ASL datasets can be seen in Table \ref{tab:externaldata}. For the IXI dataset the proposed method achieved a mean DSC of 0.958$\pm$0.006, a mean ASSD of 1.7$\pm$0.2mm, and a mean HD95 of 4.4$\pm$0.8mm, with similar results for the FSM and ASL datasets. Results from ROBEX were comparable in performance, while the performances of MONSTR and Synthstrip were numerically worse and better, respectively, keeping in mind that the ground truth in these datasets included the subarachnoid CSF. Finally, results for the CC359 dataset can be seen in Table \ref{tab:manualdata}. The proposed method achieved a mean DSC of 0.945$\pm$0.005, a mean ASSD of 3.1$\pm$0.6mm and a mean HD95 of 11.5$\pm$3.3mm. For this dataset, we also report metrics for ROBEX and Synthstrip, however, due to technical issues with MONSTR, we were not able to get that method to work on this dataset.

\begin{table}[!b]
\caption{Mean DSC, ASSD, and HD95 for the baseline methods and the proposed method for the held-out test data, comprising 547 MRIs from both the ADNI and ASAP datasets.} 
\label{tab:heldoutdata}
\begin{center}       
\begin{tabular}{lccc}
    \hline
    \multicolumn{4}{c}{Metric $\pm$ STD } \\
    \hline
    Method & DSC(\%) & ASSD(mm) & HD95(mm) \\
    \hline
    MONSTR  & 96.4 $\pm$ 1.4    & 1.4 $\pm$ 0.5  & 4.4 $\pm$ 1.2 \\
    ROBEX  & 96.9 $\pm$ 1.4  & 1.1 $\pm$ 0.5  & 4.0 $\pm$ 1.6 \\
    Synthstrip   & 96.3 $\pm$ 1.1   & 1.5 $\pm$ 0.4  & 4.6 $\pm$ 0.8 \\
    Proposed  & 96.4 $\pm$ 0.6   & 1.4 $\pm$ 0.2  & 3.6 $\pm$ 1.1 \\
    \hline
   \end{tabular}
\end{center}
\end{table}

\begin{table}[!b]
\caption{Mean DSC, ASSD, and HD95 for the baseline methods and the proposed method for the IXI, FSM and ASL datasets, comprising 50, 38, and 43 MRIs, respectively} 
\label{tab:externaldata}
\begin{center}       
\begin{tabular}{lccc}
    \hline
    \multicolumn{4}{c}{Metric $\pm$ STD} \\
    \hline
    \multicolumn{4}{c}{\textbf{IXI}} \\
    \hline
    Method & DSC(\%) & ASSD(mm) & HD95(mm) \\
    \hline
    MONSTR  & 93.5 $\pm$ 1.0    & 2.5 $\pm$ 0.4  & 7.2 $\pm$ 0.8 \\
    ROBEX  & 96.2 $\pm$ 0.8  & 1.5 $\pm$ 0.3  & 4.5 $\pm$ 0.9 \\
    Synthstrip   & 97.0 $\pm$ 0.5   & 1.2 $\pm$ 0.2  & 3.6 $\pm$ 0.6 \\
    Proposed  & 95.8 $\pm$ 0.6   & 1.7 $\pm$ 0.2  & 4.4 $\pm$ 0.8 \\
    \hline
    \multicolumn{4}{c}{\textbf{FSM}} \\
    \hline
    Method & DSC(\%) & ASSD(mm) & HD95(mm) \\
    \hline
    MONSTR  & 92.9 $\pm$ 0.8    & 2.7 $\pm$ 0.3  & 7.6 $\pm$ 0.8 \\
    ROBEX  & 96.0 $\pm$ 0.8  & 1.7 $\pm$ 0.3  & 4.5 $\pm$ 0.8 \\
    Synthstrip   & 97.8 $\pm$ 0.3   & 1.0 $\pm$ 0.1  & 2.9 $\pm$ 0.4 \\
    Proposed  & 95.6 $\pm$ 0.8   & 1.9 $\pm$ 0.3  & 4.5 $\pm$ 0.7 \\
    \hline
    \multicolumn{4}{c}{\textbf{ASL}} \\
    \hline
    Method & DSC(\%) & ASSD(mm) & HD95(mm) \\
    \hline
    MONSTR  & 84.1 $\pm$ 9.5    & 5.8 $\pm$ 3.3  & 11.7 $\pm$ 4.2 \\
    ROBEX  & 96.8 $\pm$ 1.1  & 1.7 $\pm$ 0.3  & 3.9 $\pm$ 2.2 \\
    Synthstrip   & 97.3 $\pm$ 0.5   & 1.2 $\pm$ 0.2  & 3.4 $\pm$ 0.8 \\
    Proposed  & 95.8 $\pm$ 0.7   & 1.8 $\pm$ 0.4  & 4.9 $\pm$ 2.6 \\
    \hline
   \end{tabular}
\end{center}
\end{table}

\begin{table}[!ht]
\caption{Mean DSC, ASSD, and HD95 for the ROBEX, Synthstrip and the proposed method for the manually labelled data, comprising 12 MRIs from the CC359 dataset.} 
\label{tab:manualdata}
\begin{center}       
\begin{tabular}{lccc}
    \hline
    \multicolumn{4}{c}{Metric $\pm$ STD } \\
    \hline
    Method & DSC(\%) & ASSD(mm) & HD95(mm) \\
    \hline
    ROBEX  & 95.8 $\pm$ 0.6  & 2.13 $\pm$ 0.61  & 8.8 $\pm$ 3.48 \\
    Synthstrip   & 92.1 $\pm$ 0.9   & 4.17 $\pm$ 0.69  & 13.85 $\pm$ 3.41 \\
    Proposed  & 94.5 $\pm$ 0.5   & 3.07 $\pm$ 0.60  & 11.51 $\pm$ 3.31 \\
    \hline
   \end{tabular}
\end{center}
\end{table}

\begin{figure} [!t]
    \begin{center}
    \begin{tabular}{c} 
    \includegraphics[height=4.9cm]{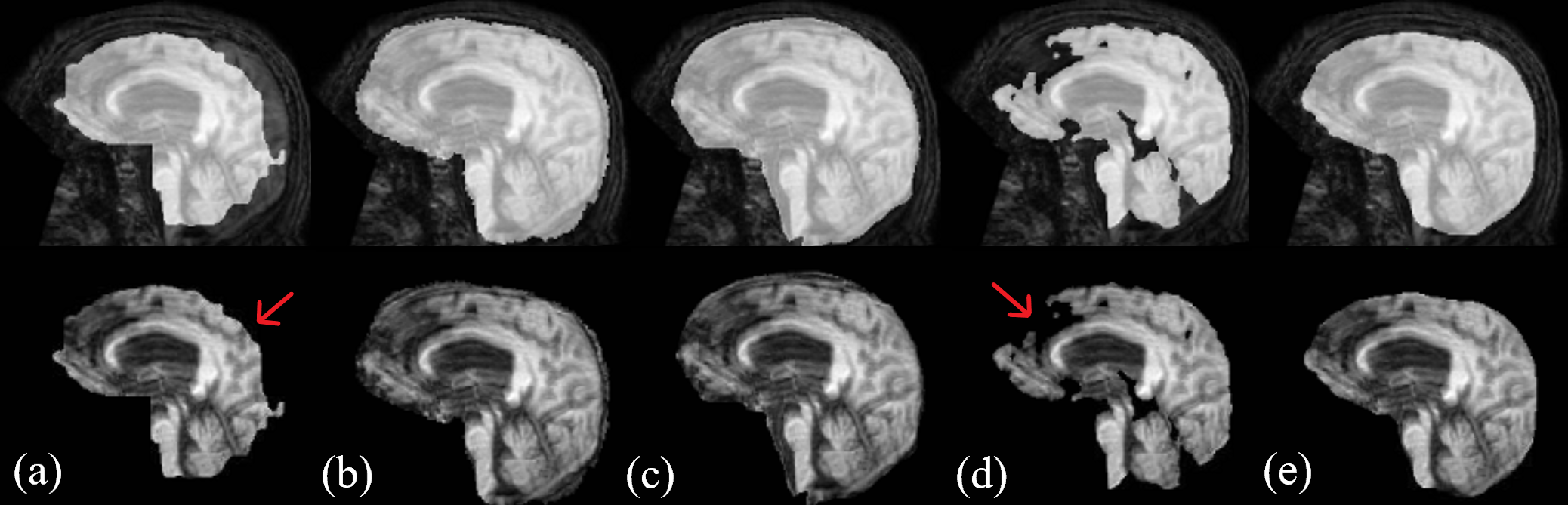}
    \end{tabular}
    \end{center}
    \caption[example] 
       { \label{fig:movement} 
Automatic brain segmentation of an image with movement artefacts. Methods are (a) MONSTR, (b) ROBEX, (c) Synthstrip, (d) Synthstrip without CSF, (e) Proposed. Red arrows indicate undersegmentations. The images in the top row display the masks overlaid onto the image, while the bottom row shows the skull stripped brain.}
   \end{figure}

\begin{figure} [!t]
    \begin{center}
    \begin{tabular}{c} 
    \includegraphics[height=4.9cm]{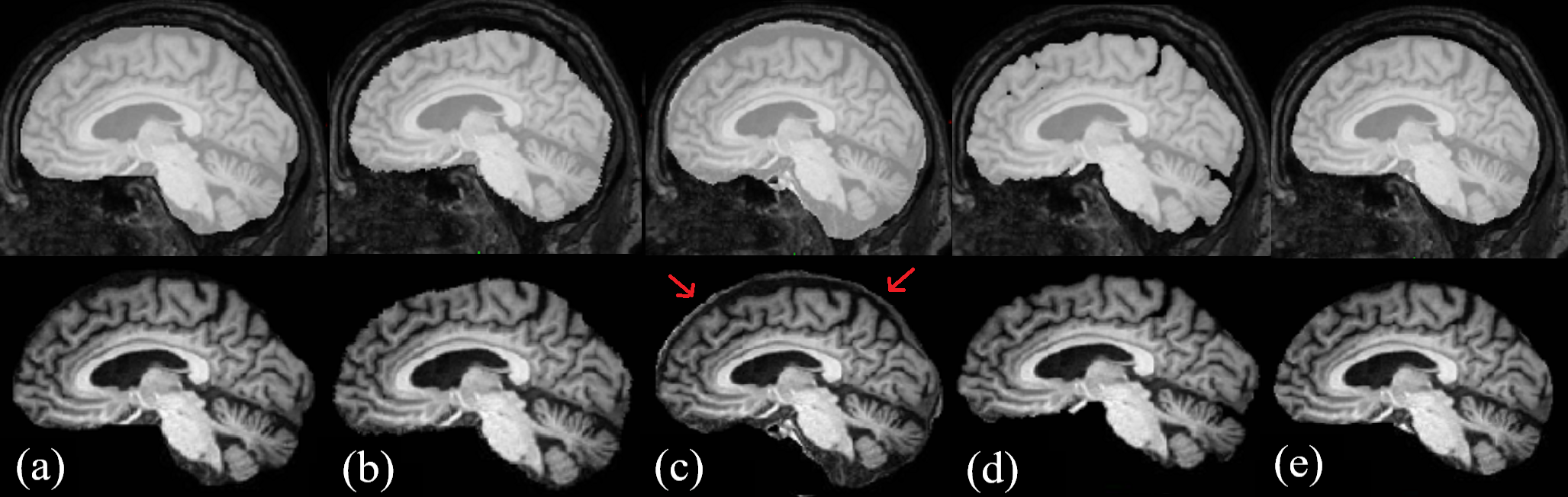}
    \end{tabular}
    \end{center}
    \caption[example] 
       { \label{fig:synthstrip} 
Automatic brain segmentation of a standard T1-weighted MRI from a healthy subject. Methods are (a) MONSTR, (b) ROBEX, (c) Synthstrip, (d) Synthstrip without CSF, (e) Proposed. Red arrows indicate oversegmentationCSF. The images in the top row display images with masks overlaid, while the bottom row shows the skull stripped brain.}
   \end{figure}

\begin{figure} [!ht]
    \begin{center}
    \begin{tabular}{c} 
    \includegraphics[height=5.3cm]{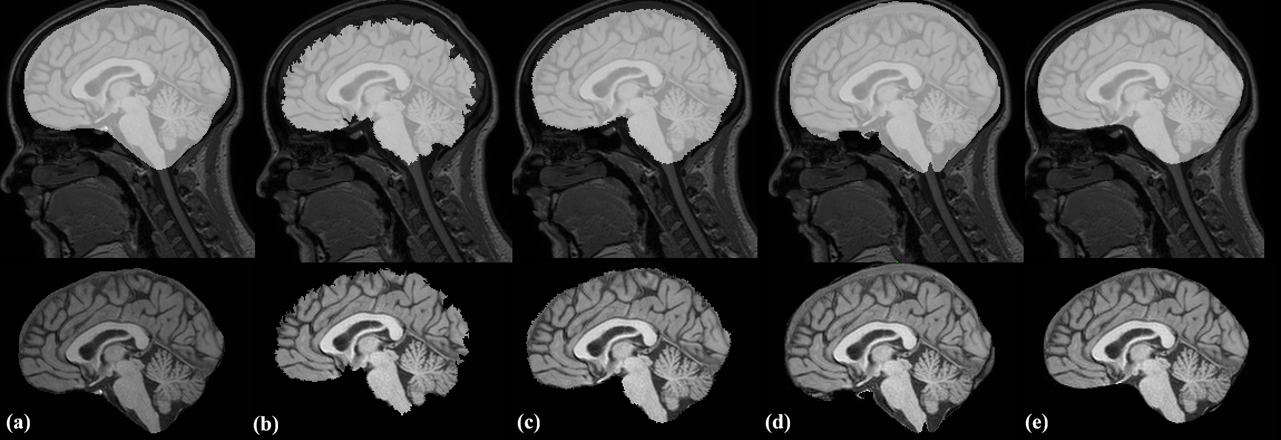}
    \end{tabular}
    \end{center}
    \caption[example] 
       { \label{fig:external} 
Automatic brain segmentation of a sample from the IXI dataset. Methods are (a) silver-standard ground truth, (b) MONSTR, (c) ROBEX, (d) Synthstrip, (e) Proposed.}
   \end{figure}

For qualitative analysis, we performed a visual inspection of results from all methods. Our analysis shows that the proposed method is highly consistent in the amount of CSF included in the segmentations. The method also demonstrates high robustness towards pathologies and movement artefacts, as demonstrated in Figure \ref{fig:movement}, where MONSTR and Synthstrip (without CSF) remove a significant amount of brain tissue, while the proposed method better preserves brain tissue. In some cases, MONSTR removed a large amount of tissue in the brainstem, which is often critical to subsequent analysis, while our model correctly preserved it. Figure \ref{fig:synthstrip} shows an example from the held-out test set, showcasing Synthstrip's oversegmentation, where it occasionally includes the meninges around the brain tissue. This inconsistent inclusion of the meninges can be problematic for subsequent image processing. An example from the external indepdendent test set can be seen in Figure \ref{fig:external}. 

The proposed method produced visually accurate segmentations in 99.3\% of cases (out of 547) in the held-out test set, and 99.2\% of cases (out of 131) in the external independent set. In the rare case of failure, the boundary of the segmentation mask in the failed images was accurate around the brain; however, the model oversegmented into high-intensity areas around the brain, such as the fat tissue in the neck.
Finally, the proposed model is computationally efficient; with an average inference time of less than 20 seconds per image, which is similar to other machine learning-based methods.

\newpage
\section{CONCLUSION}
In conclusion, we present a novel deep learning-based brain extraction method for T1-weighted MRI. Our method leverages a 3D U-net, trained on downsampled images without a significant loss of detail. Training the model on downsampled images enables highly efficient computation, with the possibility of using larger batch sizes. We also propose a novel loss function that performs well on single-object segmentation tasks without the need for high-quality ground truth data. Our quantitative results demonstrate that our model produces highly consistent results with low standard deviations in all metrics. Visually, our model demonstrates strong performance in consistently preserving the outer surface of the brain, a task that is considered particularly challenging, especially in the presence of pathology. Brain tissue is rarely misclassified as non-brain tissue, although non-brain tissue is occasionally misclassified as brain tissue. Finally, the model generalizes well on unseen data, with both qualitative and quantitative analyses indicating high-quality segmentations free from noticeable errors. Future improvements include training with more diverse data and more advanced post-processing approaches. The method can be accessed at https://github.com/HjaltiThrastarson/skullStrip25.


\newpage

\acknowledgments 
 
This work was supported by the Icelandic Centre for Research (RANNIS) under grant 200101-5601 and the University of Iceland Research Fund.

\noindent *The ASAP Neuroimaging Initiative: University of Cologne, Germany: Thilo van Eimeren, thilo.van-eimeren@uk-koeln.de, Kathrin Giehl, kathrin.giehl@uk-kolen.de, Elena Doering, elena.doering@uk-koeln.de; Turku University Hospital, Finland: Valtteri Kaasinen, valtteri.kaasinen@tyks.fi; National Research Council, Italy: Andrea Quattrone, an.quattrone@unicz.it; University of Campania, "Luigi Vanvitelli”, Italy: Alessandro Tessitore, alessandro.tessitore@unicampania.it; Universidad de Navarra, Spain: María Rodríguez Oroz, mcroroz@unav.es; University of Florida, USA: David Vaillancourt, vcourt@ufl.edu; FLENI Foundation Buenos Aires, Argentina: Julieta Arena, jarena@fleni.org.ar; University of Salerno, Italy: Marina Picillo, mpicillo@unisa.it, Paolo Barone, pbarone@unisa.it, Maria Teresa Pellecchia, m.pellecchia@unisa.it.

\bibliography{report} 
\bibliographystyle{spiebib} 

\end{document}